\documentclass[aps,pra,reprint,showpacs,amssymb,amsmath,longbibliography]{revtex4-1}
\usepackage{graphicx}

\def\be{\begin{equation}}
\def\ee{\end{equation}}
\def\ber{\begin{eqnarray}}
\def\eer{\end{eqnarray}}
\def\bern{\begin{eqnarray*}}
\def\eern{\end{eqnarray*}}

\def\rv{\mathbf{r}}

\def\kv{\mathbf{k}}

\def\0v{\mathbf{0}}
\def\1v{\mathbf{1}}
\def\2v{\mathbf{2}}
\def\3v{\mathbf{3}}

\def\pa{\partial}

\begin{document}

\title{Quasi-low-dimensional electron gas with one populated band as a testing ground for time-dependent density-functional theory}
\author{Vladimir~U.~Nazarov}
\affiliation{Research Center for Applied Sciences, Academia Sinica, Taipei 11529, Taiwan}

\begin{abstract}
We  find the analytical solution to the time-dependent density-functional theory (TDDFT) problem
for the  quasi-low-dimensional (2D and 1D) electron gas (QLDEG)
with  only one band filled in the  direction perpendicular to the system extent.
The theory is developed at the level of TD exact exchange and yields the exchange potential as an explicit nonlocal operator of 
the spin-density. 
The dressed interband (image states) excitation spectra of the Q2DEG are calculated,
while the comparison with the Kohn-Sham (KS)  transitions provides  the insight into  the qualitative and quantitative role
of the many-body interactions. Important cancellations between the Hartree $f_H$ and the exchange $f_x$ kernels
are found in the low-density limit, shedding light on the interrelations between the KS and many-body excitations.
\end{abstract}


\maketitle

Density-functional theory (DFT) \cite{Kohn-65} and its time-dependent counterpart TDDFT \cite{Runge-84} are presently, by far, the most popular methods to conceive the ground-state and excitation, respectively, properties of atomic, molecular, and condensed matter systems.
Both DFT and TDDFT require the knowledge of the exchange-correlation (xc) potentials, $v_{xc}(\rv)$ and $v_{xc}(\rv,t)$, respectively. 
Although the exact xc  potentials exist in principle, they are never known  for non-trivial systems, 
making us  resort to approximations.

The  potentials now overwhelmingly used in applications are  local functions of the electron density or also of  its spatial derivatives,
the local-density approximation (LDA) \cite{Kohn-65} and the generalized-gradient approximation (GGA) \cite{Perdew-96}, respectively.
While very simple and efficient in implementations, these approximations suffer from well known  deficiencies. 
The one of our concern here will be the inherent dimensionality dependence
of both LDA and GGA, i.e., their having distinct 3D, 2D, and 1D versions,
which makes them unreliable and even poorly substantiated in the case of the systems of 
intermediate dimensionality, such as quasi-low-dimensional materials. A truly first-principles xc functional, 
being one and the same for all systems, must work equally well for different dimensionalities, including the intermediate ones.
The exact exchange (EXX) [or optimized-effective potential (OEP)] \cite{Sharp-53,Talman-76} stands out in DFT as a first-principles 
potential not, in particular, bound to any specific dimensionality. This potential obeys a number of important requirements of the exact
theory, such as the correct asymptotic behavior  $-e^2/r$ for finite systems, the support of image states at surfaces and in low-dimensions \cite{Horowitz-06,Engel-14,Nazarov-16-2}, it produces \cite{Grabo-97,Mori-Sanchez-06} the derivative discontinuity  in the energy dependence
on the fractional electrons number \cite{Perdew-82}, and it is free from self-interaction. The time-dependent version of the EXX theory has been 
developed \cite{Ullrich-95,Gorling-97} and found to support the excitonic effect in semiconductors \cite{Kim-02}.
For all the advantages, an unfortunate drawback of the EXX theory is the extreme complexity of its implementation. It is the orbital-dependent formalism
which involves the solution of the notoriously tedious OEP integral equation \cite{Sharp-53,Talman-76}.
This  has prevented EXX from becoming widely used in applications, and even qualitative insights are often obscured by heavy numerical
difficulties. 

It, therefore, came recently as a surprise  that for the quasi-low-dimensional electron gas (QLDEG) with only  one band   populated
in the transverse direction,
the ground-state EXX problem has a simple explicit solution in terms of the (spin-) density \cite{Nazarov-16-2}.
\begin{figure}[h!]
\includegraphics[width= 0.7 \columnwidth, trim= 0 30 150 20, clip=true]{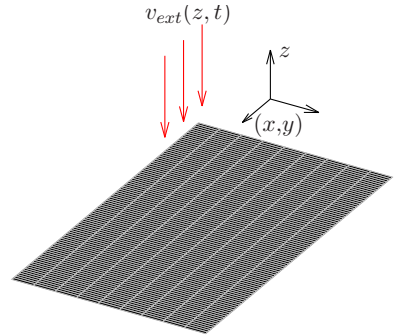}
\caption{\label{syse} (color online)
Schematics of the Q2DEG under the action of a time-dependent external potential.}
\end{figure}
A natural question arises whether the same route can be taken to build the analytical EXX theory of many-body excitations in QLDEG.
In this Letter we give to this a positive answer  by finding an explicit solution to the TD exchange potential in terms of the TD spin-density for QLDEG with one band populated.
For the solution to be expressible through the density, the applied perturbation must not change the symmetry
of the QLDEG, as is discussed below.

We start from the ground-state of a Q2DEG (for 1D case, see below), 
uniform in the $xy$-plane and confined in the $z$ direction by a potential
$v_{ext}(z)$. The in-plane and the perpendicular variables separate in this case. 
We further assume that only the states $\mu^\uparrow_0(z)$  and $\mu^\downarrow_0(z)$, one for each spin orientation,
are occupied in the $z$-direction \cite{Nazarov-16-2}, leading to the
electrons' wave-functions of the form
\begin{equation}
\psi^\sigma_{\kv_\|}(\rv)= \frac{1}{\sqrt{\Omega}} e^{i \kv_\| \cdot \rv_\|} \mu^\sigma_0(z),
\end{equation}
where $\Omega$ is the normalization area
To this system we apply a TD potential, which is  assumed to depend on the
$z$ coordinate only (see Fig.~\ref{syse}) and, by this, it  preserves the system's lateral  uniformity during the time-evolution.
We will see that a wealth of many-body phenomena are preserved within these constraints, while the gain is
the system admitting an analytical solution.

The main result of this Letter is that, with the above setup, the TDEXX potential  is
\begin{equation}
\begin{split}
  v^\sigma_x(z,t)   = -\frac{1}{n^\sigma_{2D}}
    \int    \frac{F_2(k_F^{\sigma}|z -  z'|)}{|z  - z'|}  n^\sigma(z',t) d z' ,
\end{split}
\label{main152}
\end{equation} 
where $ n^\sigma(z,t)$ is the spin-density,
\begin{equation*}
F_2(x)=1+\frac{ L_1(2 x)-I_1(2 x)}{x},
\label{F2D}
\end{equation*}
$L_1$ and $I_1$ are the first-order modified Struve and Bessel functions \cite{Prudnikov,Mathematica}, respectively, 
$n^\sigma_{2D}=\int_{-\infty}^\infty n^\sigma(z,t) d z$ is the 2D spin-density, which does not change during the time-evolution,
and $k_F^{\sigma}=\sqrt{4\pi n^\sigma_{2D}}$ is the corresponding 2D Fermi radius.
We derive Eq.~(\ref{main152}) in Appendix \ref{AA}  with the use of the adiabatic-connection method  \cite{Gorling-94,Gorling-97}.
In the linear-response regime, Eq.~(\ref{main152}) gives immediately for the exchange kernel
\begin{equation}
f^{\sigma\sigma'}_x(z,z',\omega)=\frac{\delta v_x^\sigma(z,\omega)}{\delta n^{\sigma'}(z',\omega)}=
-\frac{1}{n^\sigma_{2D}} \frac{F_2(k_F^{\sigma}|z -  z'|)}{|z  - z'|}\delta_{\sigma \sigma'}.
\label{fx} 
\end{equation}
Notably, $f_x$ of Eq.~(\ref{fx}) is frequency-independent. We, however, emphasize 
that Eqs.~(\ref{main152}) and (\ref{fx}) are by no means an adiabatic approximation: Our detailed derivation
in Appendix \ref{AA} shows that they hold {\it exactly} within the fully dynamic TDEXX for QLDEG with one band filled,
provided that the exciting field is applied perpendicularly to the layer.

\begin{figure}[h!]
\includegraphics[width=  \columnwidth, clip=true, trim=65 4 1 1]{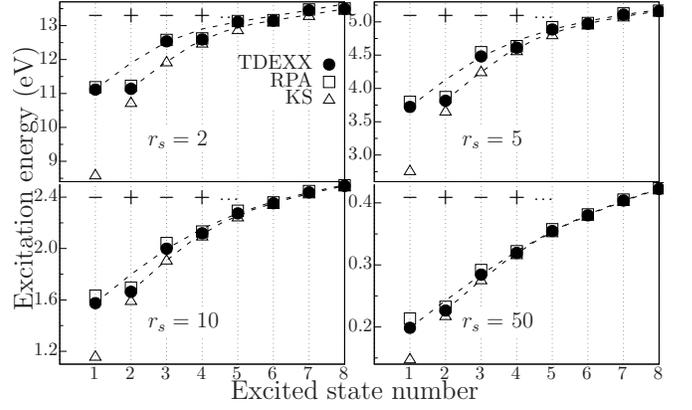}
\caption{\label{eig22}
Excitation energies of a spin-neutral Q2DEG with one transverse band filled.
Circles, squares, and triangles are TDEXX, RPA, and KS excitation energies, respectively. 
Plus and minus signs mark even and odd excitations, respectively. 
Dashed lines connect the energies of the  even and odd self-oscillations, separately.}
\end{figure}

\begin{figure}[h!]
\includegraphics[width=  \columnwidth, clip=true, trim=65 4 1 1]{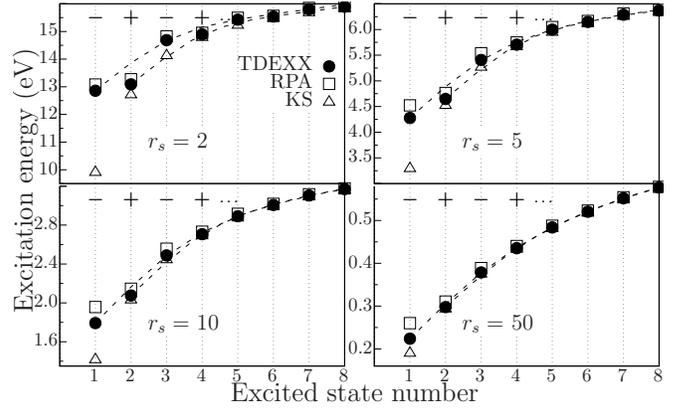}
\caption{\label{eig22pp}
The same as Fig.~\ref{eig22}, but for the fully spin-polarized Q2DEG.}
\end{figure}

We use the kernel of Eq.~(\ref{fx}) with the basic linear-response TDDFT equality \cite{Gross-85,Giuliani&Vignale}
\begin{equation}
\begin{split}
\left(\chi^{-1}\right)^{\sigma\sigma'} (z,z',\omega)  &=  
\left(\chi^{-1}_s\right)^{\sigma\sigma'}   (z,z',\omega) - f_H(z,z') \\
&- f^{\sigma\sigma'}_x(z,z',\omega),
\end{split}
\label{chchs}
\end{equation}
where $\chi$ and $\chi_s$ are the interacting-electrons and Kohn-Sham (KS) spin-density-response functions, respectively,
the latter given in our case by
\begin{equation}
\begin{split}
& \chi_s^{\sigma\sigma'}(z,z',\omega) =  n^\sigma_{2D} \mu^\sigma_0(z) \mu^{\sigma}_0(z') 
\sum\limits_{n=1}^\infty 
\left(
\frac{1}{ \omega+\lambda_0^\sigma - \lambda_n^\sigma+i 0_+} \right. \\ & \left.  -
\frac{1}{ \omega-\lambda_0^\sigma + \lambda_n^\sigma+i 0_+}
\right)
\mu^{\sigma}_n(z)
 \mu^\sigma_n(z') \delta_{\sigma\sigma'},
 \end{split} 
\label{chs1}
\end{equation}
where $\lambda^\sigma_n$ and $\mu^{\sigma}_n(z)$ are the eigenenergies and the eigenfunctions of the perpendicular motion, respectively.
A remarkable property of the KS response function of our system is that it is immediately invertible to (see Appendix \ref{BB})
\begin{equation}
\begin{split}
 (\chi^{-1}_s)^{\sigma\sigma'}  (z,z',\omega)   &=   \frac{\delta_{\sigma\sigma'}}{2 n^\sigma \mu^\sigma_0(z) \mu^{\sigma}_0(z')} \\
&\times \left[ \omega^2 X_1^\sigma(z,z') - X_2^\sigma(z,z') \right] ,
\end{split}
\label{chsm}
\end{equation}
with
\begin{equation}
\begin{split}
 X_1^\sigma(z,z')  &=  
\sum\limits_{n=1}^\infty 
\frac{\mu^{\sigma}_n(z) \mu^\sigma_n(z')}{\lambda_n^\sigma-\lambda_0^\sigma} \\
&=\left(\hat{h}^\sigma_s-\lambda_0^\sigma\right)^{\! -1} \left[\delta(z-z')-\mu^{\sigma}_0(z) \mu^\sigma_0(z')\right],
\end{split}
\label{X1}
\end{equation}
\begin{equation}
\begin{split}
X_2^\sigma(z,z') &=
\sum\limits_{n=1}^\infty \!
(\lambda_n^\sigma - \lambda_0^\sigma)
\mu^{\sigma}_n(z) \mu^\sigma_n(z') \\
&=\left(\hat{h}^\sigma_s-\lambda_0^\sigma\right) \delta(z-z'),
\end{split}
\label{X2}
\end{equation}
where $\hat{h}^\sigma_s$ is the static KS Hamiltonian
\footnote{The operator $\hat{h}^\sigma_s-\lambda_0^\sigma$ is invertible on the subspace of functions orthogonal to $\mu_0^\sigma(z)$,
to which the function in the brackets on the right-hand side of Eq.~(\ref{X1}) belongs.}.
The Hartree part of the kernel is
\begin{equation}
f^{\sigma\sigma'}_H(z,z')  = - 2 \pi |z-z'|.  
\label{fH}
\end{equation}
The many-body excitation energies $\omega$ are found from the equation
\begin{equation}
\sum\limits_{\sigma'} \int \left(\chi^{-1}\right) ^{\sigma\sigma'}(z,z',\omega) \delta n^{\sigma'}(z',\omega) d z' =0,
\label{Eq0}
\end{equation}
where $\delta n^\sigma(z)$ is the self-oscillation of the spin-density

With the use of Eqs.~(\ref{chchs}) and (\ref{chsm})-(\ref{X2}), Eq.~(\ref{Eq0}) can be rewritten as the following 
eigenvalue problem 
\begin{equation}
\begin{split}
 &\left( \hat{h}_s^\sigma    -  \lambda_0^\sigma \right)   \left[ 
 \left( \hat{h}_s^\sigma  -  \lambda_0^\sigma \right) \! y^\sigma(z) + 2  n^\sigma_{2D}  \right. \\ & \times \! \! \left.
     \int  \mu^\sigma_0(z)   \sum_{\sigma'}  f_H^{\sigma\sigma'}(z,z')  \mu^{\sigma'}_0(z') y^{\sigma'}(z') d z' + 2  n^\sigma_{2D} 
 \right. \\ &\times  \! \! \left. 
  \int \mu^\sigma_0(z) f_x^{\sigma\sigma}(z,z') \mu^\sigma_0(z') y^\sigma(z') d z' \right]
 =  \omega^2   y^\sigma(z),
\end{split}
\label{ME}
\end{equation}
where  $y^\sigma(z)= \delta n^\sigma(z)/\mu^\sigma_0(z)$.

We have found the eigenvalues and eigenfunctions of Eq.~(\ref{ME}) 
numerically  on a $z$-axis grid for a number of the EG densities.
The confining potential $v_{ext}(z)$ was chosen that of the 2D positive charge background.
The static KS problem was solved self-consistently with the use of the EXX potential, which
is that of Eq.~(\ref{main152})
with the ground-state density in place of the TD one \cite{Nazarov-16-2}.
Results for the eigenenergies of the excited states are presented in Figs.~\ref{eig22} and \ref{eig22pp},
for spin-neutral and fully spin-polarized Q2DEG, respectively, where TDEXX is compared to the random-phase approximation (RPA)
[setting $f_x=0$ in Eq.~(\ref{ME})] and with the KS transitions [setting $f_x=f_H=0$  in Eq.~(\ref{ME})]. Obviously, the first excited state is influenced strongly 
by the many-body interactions, resulting in the both TDEXX and RPA being very different from the single-particle KS transition.
This effect, however, weakens for higher excited states. 
Secondly, the difference between the TDEXX and RPA increases with the growth of $r_s$ (decrease of the density),
the former moving to the KS values,
which is more pronounced for the spin-polarized than for the spin-neutral  EG. This has an elegant explanation:
Expanding Eq.~(\ref{fx}) in powers of $k^\sigma_F$, we can write at small $k^\sigma_F$
\begin{equation}
f^{\sigma\sigma'}_x(z,z',\omega)\approx
\left[ - \frac{32}{3 k_F^\sigma} + 2 \pi |z-z'| \right] \delta_{\sigma\sigma'}.
\label{fxexp} 
\end{equation}
\begin{figure}[h!]
\includegraphics[width=  \columnwidth, clip=true, trim=40 0 10 0]{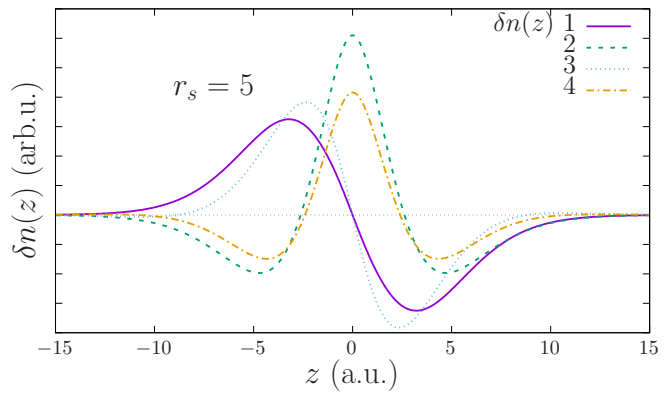}
\caption{\label{vr} (color online)
Self-oscillations of the density of the spin-neutral EG of $r_s=5$, corresponding to the transitions to the first four excited states.}
\end{figure}
\noindent Noting that the first term in Eq.~(\ref{fxexp}) is a constant and, consequently, it does not play a role in $f_x$, and  comparing with Eq.~(\ref{fH}), we conclude that, for a dilute EG, the exchange part of the kernel by {\it a half and fully cancels the Hartree part, for the spin-neutral and fully spin-polarized EG, respectively}. In the fully spin-polarized case, at low densities, this brings the many-body excitation energies back to the KS values, as can be observed in Fig.~\ref{eig22pp}.

In contrast to the KS transitions, TDEXX and RPA excitation energies split into the even and odd series, 
the values changing  smoothly within each, while jumping across the series.  
In Figs.~\ref{eig22} and \ref{eig22pp}, the points within each series are connected with  dashed lines serving as  eye-guides.
In Fig.~\ref{vr}, we plot the even and odd self-oscillations $\delta n(z)$ themselves.

The quasi-1D electron gas admits the same treatment  as the Q2DEG above, leading to the following results
(cf. the static case \cite{Nazarov-16-2}). For the exchange kernel we have
\begin{equation}
f^{\sigma\sigma'}_x(\boldsymbol{\rho},\boldsymbol{\rho}',\omega)=
-\frac{1}{n^\sigma_{1D}} \frac{F_1(k_F^{\sigma}|\boldsymbol{\rho} -  \boldsymbol{\rho}'|)}{|\boldsymbol{\rho}  - \boldsymbol{\rho}'|}\delta_{\sigma \sigma'},
\label{fx1} 
\end{equation}
where $\boldsymbol{\rho}=(x,y)$, the wire is stretched along the $z$-axis,
\begin{equation}
F_1(x)=  \frac{1}{2\pi}  G_{2,4}^{2,2}\left[x^2  \left|
\begin{array}{l}
 \frac{1}{2},1 \\[0.25 em]
 \frac{1}{2},\frac{1}{2},-\frac{1}{2},0 \\
\end{array} \right.
\right],
\label{F1D}
\end{equation}
and
$G_{p,q}^{m,n}\left[x  \left|
\begin{array}{lll}
 a_1, &..., &a_p \\
 b_1, &..., &b_q \\
\end{array} \right.
\right]$ is the Meijer G-function \cite{Prudnikov,Mathematica}. The Hartree kernel is
\begin{equation}
f^{\sigma\sigma'}_H(\boldsymbol{\rho},\boldsymbol{\rho}')= -2 \log \left( k_F^\sigma |\boldsymbol{\rho}  - \boldsymbol{\rho}'| \right).
\end{equation}

As shown earlier \cite{Nazarov-16-2}, 
the assumption of the QLDEG having one spin-state occupied in the perpendicular direction  is not very restrictive: 
This is a regime actually realizing at $r_s>1.46$ and $r_s>0.72$, for the Q2D and Q1D cases, respectively,
provided the confining potential is that of the positive 2D (1D) uniform background.
The second feature of our setup,  that of the perturbation field being applied perpendicularly to the layer, is important:
By this we do not study the excitation spectra of the 2D (1D)  EG proper, 
which problem  has been extensively addressed in the literature before \cite{Stern-67,Giuliani&Vignale},
but we are concerned with the interband excitations, which are the excitations to the image states of QLDEG.
The latter excitations we handle as dressed, i.e., accounting for the many-body dynamic interactions, doing this at the level of TDEXX.
With the understanding of the above, our theory is exact.

The localized Hartree-Fock potential (LHF) \cite{Sala-01} has recently attracted new attention 
as a single-particle potential providing the best possible fulfillment of the many-body TD Schr\"{o}dinger equation
by a Slater determinant wave-function \cite{Nazarov-13-2,Nazarov-15-2} and, in the spirit
of the ``direct-energy'' potentials \cite{Levy-14}, yielding the energy as a sum of KS eigenvalues.
It has been recently shown \cite{Nazarov-16-2} that for QLDEG with one populated band in its ground-state, LHF potential 
coincides exactly up to a constant with the EXX one.  The same is true in the TD case, the proof of which
is a repetition of that given in  Sec.~V of Ref.~\cite{Nazarov-16-2} for the static case, 
with all the functions acquiring an additional time-argument $t$.

In conclusions, 
we have identified the quasi-low-dimensional electron gas with one occupied band  as a unique system admitting analytical
or semi-analytical solution of the many-body excitation problem by means of the time-dependent density-functional theory
at the level of the time-dependent exact-exchange. The fundamental quantities of TDDFT, such as the time-dependent
exchange potential and the exchange kernel, have been constructed as an explicit nonlocal operator of the spin-density
and purely analytically, respectively. We have applied our theory to obtain the  interband excitation spectra  (excitation to image states) of Q2DEG.
The low-lying excited states are shown to be strongly affected by the many-body interactions for the EG of higher densities.
In the low-density regime, we have shown that  the exchange kernel cancels the Hartree  one  by a half and entirely, 
in the case of the spin-neutral and fully spin-polarized EG, respectively. 
This  demonstrates how qualitatively wrong and inconsistent the often used  random-phase approximation (i.e., the account of the Hartree part of the kernel only) may be.
For the dilute fully spin-polarized QLDEG this leads to an important conclusion that the Kohn-Sham excitation energies can be, at the same time, the true excitation energies of a many-body system. 

We, finally, argue that QLDEG with one populated band  has a promise to be extendable to yield analytical  or semi-analytical results with inclusion of correlations, further enriching our understanding of DFT and TDDFT in mesoscopic physics.


%

\onecolumngrid
\appendix

\section{Time-dependent exchange potential [Proof of Eq.~(\ref{main152})]}
\label{AA}

We follow the adiabatic connection method \cite{Gorling-94,Gorling-97}.
The adiabatic connection Hamiltonian is (for brevity, in the following we omit the spin index)
\begin{equation}
\hat{H}_\lambda(t) = \sum\limits_i \left[ -\frac{1}{2} \Delta_i +v_{ext}(\rv_i,t) + \tilde{v}_\lambda(\rv_i,t) \right] 
+ \sum\limits_{i<j} \frac{\lambda}{|\rv_i-\rv_j|}.
\label{Hl}
\end{equation}
The corresponding density-matrix satisfies the Liouville's equation
\begin{equation}
i \frac{\pa \hat{\rho}_\lambda(t)}{\pa t}= [\hat{H}_\lambda(t),\hat{\rho}_\lambda(t)].
\label{l}
\end{equation}
To the first order in $\lambda$ we have
\begin{align}
&i \frac{\pa \hat{\rho}_0(t)}{\pa t}= [\hat{H}_0(t),\hat{\rho}_0(t)],\\
&i \frac{\pa \hat{\rho}_1(t)}{\pa t}= [\hat{H}_0(t),\hat{\rho}_1(t)]+[\hat{H}_1(t),\hat{\rho}_0(t)],
\label{l1}
\end{align}
where
\begin{align}
&\hat{H}_\lambda(t) =\hat{H}_0(t)+ \lambda \hat{H}_1(t),\\
&\hat{\rho}_\lambda(t) =\hat{\rho}_0(t)+ \lambda \hat{\rho}_1(t),\\
&\hat{H}_0(t) = \sum\limits_i \left[ -\frac{1}{2} \Delta_i +v_{ext}(\rv_i,t) + \tilde{v}_0(\rv_i,t) \right],\\
&\hat{H}_1(t) = \sum\limits_i \tilde{v}_1(\rv_i,t) 
+ \sum\limits_{i<j} \frac{1}{|\rv_i-\rv_j|},\label{H1}\\
&\tilde{v}_\lambda(\rv,t)=\tilde{v}_0(\rv,t)+ \lambda \tilde{v}_1(\rv,t).
\label{v01}
\end{align}

Let for $t\le 0$ the external potential be time-independent and the system be in its ground state
with the KS (Slater-determinant) wave-function $|0\rangle$, where $|\alpha\rangle$ is the orthonormal complete set of the KS
eigenfunctions of the Hamiltonian $\hat{H}_0(0)$. Let at $t=0$ the time-dependent part of the external potential switch on.
Then,  $|\alpha(t)\rangle$, which satisfies
\begin{align}
&i \frac{\pa |\alpha(t)\rangle}{\pa t}= \hat{H}_0(t) |\alpha(t)\rangle,
\label{ts}\\
&|\alpha(0)\rangle=|\alpha\rangle,
\end{align}
is also the orthonormal complete set of the Slater-determinant wave-functions at each particular time $t$.
Taking matrix elements of Eq.~(\ref{l1}), we write with the help of Eq.~(\ref{ts})
\begin{equation}
i \langle \alpha(t)|\frac{\pa \hat{\rho}_1(t)}{\pa t}|  \beta(t) \rangle = 
\langle i \frac{\pa \alpha(t)}{\pa t} | \hat{\rho}_1(t)| \beta(t)\rangle -
\langle \alpha(t)| \hat{\rho}_1(t)| i \frac{\pa \beta(t)}{\pa t} \rangle
+(\delta_{\beta 0}-\delta_{\alpha 0}) \langle \alpha(t)| \hat{H}_1(t)| \beta(t)\rangle
\label{ll}
\end{equation}
or
\begin{equation}
i \frac{\pa}{\pa t} \langle \alpha(t)| \hat{\rho}_1(t)|  \beta(t) \rangle = 
(\delta_{\beta 0}-\delta_{\alpha 0}) \langle \alpha(t)| \hat{H}_1(t)| \beta(t)\rangle.
\label{lll}
\end{equation}
Besides Eq.~(\ref{lll}), we need the initial condition at $t=0$. This is obtained from Eq.~(\ref{l1}), which gives at $t\le 0$
\begin{equation}
[\hat{H}_0,\hat{\rho}_1]+[\hat{H}_1,\hat{\rho}_0]=0,
\label{l10}
\end{equation}
and which, solved with respect to $\hat{\rho}_1$, produces
\begin{equation}
\langle \alpha| \hat{\rho}_1|  \beta \rangle =
\frac{\delta_{\beta 0}-\delta_{\alpha 0}}{E_\beta-E_\alpha} \langle \alpha| \hat{H}_1|  \beta \rangle,
\label{s}
\end{equation}
where  $E_\alpha$ are the eigenvalues of $\hat{H}_0$. Together Eqs.~(\ref{lll}) and (\ref{s}) give
\begin{equation}
\langle \alpha(t)| \hat{\rho}_1(t)|  \beta(t) \rangle = 
(\delta_{\beta 0}-\delta_{\alpha 0}) \left[ \frac{\langle \alpha| \hat{H}_1|  \beta \rangle}{E_\beta-E_\alpha} -i \int_0^t \langle \alpha(t')| \hat{H}_1(t')| \beta(t')\rangle d t'
    \right].
\label{mat}
\end{equation}

Further, we calculate the TD density to the 1st order in $\lambda$
\begin{equation}
n_1(\rv,t)= {\rm Sp} \left\{ \hat{\rho}_1(t) \sum\limits_i \delta(\rv-\rv_i)   \right\}
\end{equation}
or written through the matrix elements and with account of the identity of electrons
\begin{equation}
n_1(\rv,t)= N \sum\limits_{\alpha \beta}
\left\langle \alpha(t) \left| \hat{\rho}_1(t)\right|\beta(t)\right\rangle \left\langle \beta(t)\left| \delta(\rv-\rv_1) \right| \alpha(t) \right\rangle. 
\label{n}
\end{equation}
The following facts will play a critical role below:
\begin{enumerate}
\item
\label{1}
The density operator is a single-particle operator and, therefore,
only the determinants $|\alpha(t)\rangle$ and $|\beta(t)\rangle$
which differ  by one orbital at most contribute to Eq.~(\ref{n});
\item
\label{2}
Because of Eq.~(\ref{mat}), only the elements $\langle 0(t) \left| \hat{\rho}_1(t)\right|\alpha(t) \rangle$
and $\langle \alpha(t) \left| \hat{\rho}_1(t)\right|0(t) \rangle$, $\alpha\ne 0$, are non-zero;
\item
\label{3}
Due to the symmetry of our system and the external potential varying in the $z$ direction only,
the density $n_1(\rv,t)$ can 
be a function of the $z$ coordinate only. 
We can, therefore, average Eq.~(\ref{n}) in $(x,y)$ over the normalization area $\Omega$ without changing this equation.
\end{enumerate}
Then, with account of the facts \ref{2} and \ref{3},
\begin{equation}
n_1(z,t)= 2 N \, {\rm Re} \sum\limits_{\alpha\ne 0}
\left\langle \alpha(t) \left| \hat{\rho}_1(t)\right|0(t)\right\rangle \left\langle 0(t)\left| \delta(z-z_1) \right| \alpha(t) \right\rangle, 
\label{nza0}
\end{equation}
Since
during the time-evolution of the KS system
the orbitals remain of the form
\begin{equation}
\phi_i(\rv,t)= \mu_{n_i}(z,t) \frac{e^{i \kv_{i \|} \cdot \rv_{i \|}}}{\Omega^{1/2}},
\end{equation}
and with account of the facts \ref{1} and \ref{2}, we conclude that only the matrix elements $\langle \alpha_{ns}(t) \left| \hat{\rho}_1(t)\right|0(t) \rangle=
\langle 0(t) \left| \hat{\rho}_1(t)\right|\alpha_{ns}(t) \rangle^*$, where
\begin{equation}
|\alpha_{ns}(t)\rangle=\frac{1}{(N! \Omega^N)^{1/2}}
\left|
\begin{array}{ccc}
\mu_0(z_1,t) e^{i \kv_{1 } \cdot \rv_{1 \|}} & \hdots &\mu_0(z_N,t) e^{i \kv_{1 } \cdot \rv_{N \|}}\\
\vdots &\hdots &\vdots \\
\mu_0(z_1,t) e^{i \kv_{s-1} \cdot \rv_{1 \|}} &\hdots &\mu_0(z_N,t) e^{i \kv_{s-1} \cdot \rv_{N \|}}\\
\mu_n(z_1,t) e^{i \kv_s \cdot \rv_{1 \|}} &\hdots   &\mu_n(z_N,t) e^{i \kv_s \cdot \rv_{N \|}}\\
\mu_0(z_1,t) e^{i \kv_{s+1} \cdot \rv_{1 \|}} &\hdots  &\mu_0(z_N,t) e^{i \kv_{s+1} \cdot \rv_{N \|}}\\
\vdots &\hdots &\vdots\\
\mu_0(z_1,t) e^{i \kv_{N } \cdot \rv_{1 \|}} &\hdots &\mu_0(z_N,t) e^{i \kv_{N } \cdot \rv_{N \|}}
\end{array}
\right|,
\end{equation}
$n=1,2, ...$, $s=1...N$,
and
\begin{equation}
|0(t)\rangle=\frac{1}{(N! \Omega^N)^{1/2}}
\left|
\begin{array}{ccc}
\mu_0(z_1,t) e^{i \kv_{1 } \cdot \rv_{1 \|}} & \hdots &\mu_0(z_N,t) e^{i \kv_{1 } \cdot \rv_{N \|}}\\
\vdots &\hdots &\vdots \\
\mu_0(z_1,t) e^{i \kv_{N } \cdot \rv_{1 \|}} &\hdots &\mu_0(z_N,t) e^{i \kv_{N } \cdot \rv_{N \|}}
\end{array}
\right|,
\end{equation}
contribute to Eq.~(\ref{nza0}).

We can write by Eq.~(\ref{H1})
\begin{equation} 
 \langle \alpha_{ns}(t)| \hat{H}_1(t)| 0(t)\rangle=
\langle \alpha_{ns}(t)| N \tilde{v}_1(\rv_1,t)+\frac{N (N-1)}{2 |\rv_1-\rv_2|}| 0(t)\rangle,
\label{llll}
\end{equation}
and we evaluate straightforwardly
\begin{align}
&\langle \alpha_{ns}(t)| \tilde{v}_1(z_1,t)| 0(t)\rangle= \frac{1}{N}
\int \mu^*_n(z_1,t) \tilde{v}_1(z_1,t) \mu_0(z_1,t) d z_1,\\
&\langle \alpha_{ns}(t)| \frac{1}{|\rv_1-\rv_2|} | 0(t)\rangle= \frac{2}{\Omega^2 N(N-1)} \int
\frac{ \mu^*_n(z_1,t) \mu_0(z_1,t) |\mu_0(z_2,t)|^2}{|\rv_1-\rv_2|} 
\left[ N- \Omega e^{i \kv_s\cdot (\rv_{2 \|}-\rv_{1 \|})} \rho^*(\rv_{2\|}-\rv_{1\|})
\right] d\rv_1 d\rv_2, \label{36} \\
&\langle \alpha_{ns}(t)| \delta(z-z_1)| 0(t)\rangle= \frac{1}{N}
\mu^*_n(z,t) \mu_0(z,t),
\end{align}
where
\begin{equation}
\rho(\rv_\|)= \frac{1}{\Omega} \sum\limits_{|\kv|\le k_F} e^{i \kv \cdot \rv_\|}.
\label{38}
\end{equation}
Then, by Eqs.~(\ref{llll})-(\ref{36}), and with an integration variable substitution,
\begin{equation}
\langle \alpha_{ns}(t)| \hat{H}_1(t)| 0(t)\rangle = 
\int \mu^*_{n_\alpha}(z_1,t)  \mu_0(z_1,t)
\left\{ \tilde{v}_1(z_1,t) +
\frac{1}{\Omega } \int
\frac{  |\mu_0(z_2,t)|^2}{\sqrt{(z_1-z_2)^2+\rv_{2\|}^2}} 
\left[ N- \Omega e^{i \kv_{s_\alpha}\cdot \rv_{2 \|}} \rho^*(\rv_{2\|})
\right]  d\rv_{2\|} d z_2
\right\} d z_1.
\label{39}
\end{equation}

By virtue of Eqs.~(\ref{mat}) and (\ref{39}) we can write
\begin{equation}
\begin{split}
&\langle \alpha_{n s}(t)| \hat{\rho}_1(t)|0(t) \rangle = 
-i \int\limits_0^t d t' \int \mu^*_n(z_1,t')  \mu_0(z_1,t') 
\left\{ \tilde{v}_1(z_1,t') +
\frac{1}{\Omega } \int
\frac{  |\mu_0(z_2,t')|^2}{\sqrt{(z_1-z_2)^2+\rv_{2\|}^2}} 
\left[ N- \Omega e^{i \kv_s\cdot \rv_{2 \|}} \rho^*(\rv_{2\|})
\right]  d\rv_{2\|} d z_2
\right\} d z_1 \\
& + \frac{1}{\epsilon_0-\epsilon_{n_\alpha}}
\int \mu^*_n(z_1)  \mu_0(z_1)
\left\{ \tilde{v}_1(z_1) +
\frac{1}{\Omega } \int
\frac{  |\mu_0(z_2)|^2}{\sqrt{(z_1-z_2)^2+\rv_{2\|}^2}} 
\left[ N- \Omega e^{i \kv_s\cdot \rv_{2 \|}} \rho^*(\rv_{2\|})
\right]  d\rv_{2\|} d z_2
\right\} d z_1.
\end{split}
\label{40}
\end{equation}

\begin{equation}
n_1(z)= 2 \, {\rm Re} \sum\limits_{n=1}^\infty \sum\limits_{s=1}^N
\left\langle \alpha_{n s}(t) \left| \hat{\rho}_1(t)\right|0(t)\right\rangle 
\mu_n(z,t) \mu^*_0(z,t). 
\label{nzns}
\end{equation}
In the spirit of the adiabatic connection method, with the change of $\lambda$ from $0$ to $1$,
the TD density should remain unchanged.
Since, by Eqs.~(\ref{38}) and (\ref{40}),
\begin{equation}
\begin{split}
&\sum\limits_{s=1}^N
\langle \alpha_{n s}(t)| \hat{\rho}_1(t)|0(t) \rangle = 
-i \int\limits_0^t d t' \int \mu^*_n(z_1,t')  \mu_0(z_1,t') 
\left\{ N \tilde{v}_1(z_1,t') +
\frac{1}{\Omega } \int
\frac{  |\mu_0(z_2,t')|^2}{\sqrt{(z_1-z_2)^2+\rv_{2\|}^2}} 
\left[ N^2- \Omega^2  |\rho(\rv_{2\|})|^2)
\right]  d\rv_{2\|} d z_2
\right\} d z_1 \\
& + \frac{1}{\epsilon_0-\epsilon_{n_\alpha}}
\int \mu^*_n(z_1)  \mu_0(z_1)
\left\{ N \tilde{v}_1(z_1) +
\frac{1}{\Omega } \int
\frac{  |\mu_0(z_2)|^2}{\sqrt{(z_1-z_2)^2+\rv_{2\|}^2}} 
\left[ N^2- \Omega^2  |\rho(\rv_{2\|})|^2
\right]  d\rv_{2\|} d z_2
\right\} d z_1,
\end{split}
\end{equation}
we see that $n_1(z,t)=0$ if
\begin{equation}
  \tilde{v}_1(z_1,t) =-
\frac{1}{\Omega N} \int
\frac{  |\mu_0(z_2,t)|^2}{\sqrt{(z_1-z_2)^2+\rv_{2\|}^2}} 
\left[ N^2- \Omega^2  |\rho(\rv_{2\|})|^2)
\right]  d\rv_{2\|} d z_2.
\end{equation}

According to Eq.~(\ref{Hl}), $\tilde{v}^0(\rv,t)=v_H(\rv,t)+v_{xc}(\rv,t)$ and $\tilde{v}^1(\rv,t)=0$,
where $v_H(\rv,t)$ and $v_{xc}(\rv,t)$ are Hartree and the exchange-correlations potentials, respectively. 
To the first order in $\lambda$ (\ref{v01}) this gives 
\begin{equation}
\tilde{v}_1(\rv,t)=-v_H(\rv,t)-v_x(\rv,t),
\end{equation}
where in the notation we have taken into account that to the first order  we have, by definition, exchange only
\cite{Gorling-94,Gorling-97}. On the other hand, it is easy to see that for our system
\begin{equation}
v_H(\rv,t)=\frac{N}{\Omega } \int
\frac{  |\mu_0(z',t)|^2}{\sqrt{(z-z')^2+{\rv'}_\|^2}} d\rv'_\| d z',
\end{equation}
leading us to
\begin{equation}
v_x(z,t) =-
\frac{\Omega}{N} \int
\frac{  |\mu_0(z',t)|^2}{\sqrt{(z-z')^2+{\rv'}_\|^2}} 
 |\rho(\rv'_\|)|^2  d\rv'_\| d z',
 \label{lb1}
\end{equation}

The proof of Eq.~(\ref{main152}) by the explicit integration over $\rv'_\|$ in Eq.~(\ref{lb1}) with the account of Eq.~(\ref{38}).

\section{Kohn-Sham spin-density-response function and its inverse [Proof of Eqs.~(\ref{chsm})-(\ref{X2})]}
\label{BB}

We construct the operator
\begin{equation}
 (\chi^{-1}_s)^{\sigma\sigma'}(z,z',\omega) =  \frac{1}{n_{2D}^\sigma} 
\sum\limits_{n=1}^\infty 
\left(
\frac{1}{ \omega+\lambda_0^\sigma - \lambda_n^\sigma+i 0_+}   -
\frac{1}{ \omega-\lambda_0^\sigma + \lambda_n^\sigma+i 0_+}
\right)^{-1}
\frac{\mu^{\sigma}_n(z) \mu^\sigma_n(z')}{\mu^\sigma_0(z) \mu^{\sigma}_0(z') } \delta_{\sigma\sigma'}
\label{chsm1}
\end{equation}
and directly check that for an arbitrary function $g(z)$ such that
\begin{equation}
\int g(z) dz =0,
\end{equation}
the equality holds
\begin{equation}
\int \chi^\sigma_s(z,z'',\omega) (\chi^\sigma_s)^{-1}(z'',z',\omega) g(z') d z'' d z'=g(z),
\label{inv1}
\end{equation}
where $\chi_s$ is given by Eq.~(\ref{chs1}). In arriving at Eq.~(\ref{inv1}) we have used the completeness relation
\begin{equation}
\sum\limits_{n=0}^\infty \mu_n^\sigma(z) \mu_n^\sigma(z') = \delta(z-z').
\label{compl}
\end{equation}

On the other hand, the operator $(\chi^\sigma_s)^{-1} \chi^\sigma_s$ is defined on any function $h(z)$ of the Hilbert space, and
\begin{equation}
\int (\chi^\sigma_s)^{-1}(z,z'',\omega) \chi^\sigma_s(z'',z',\omega) h(z') d z'' d z'=h(z)  - \int [\mu^\sigma_0(z')]^2 h(z') dz',
\label{inv2}
\end{equation}
where the second term on the right-hand side is a constant. Equations (\ref{inv1}) and (\ref{inv2}) prove that
$\chi_s$ of Eq.~(\ref{chs1}) and $\chi_s^{-1}$ of Eq.~(\ref{chsm1}) are inverse to each other in the sense as the density-response function
and its inverse should be.

From Eq.~(\ref{chs1}) we arrive at Eqs.~(\ref{chsm})-(\ref{X2}) by simple algebraic manipulations with the expression in the parentheses
and by the use of the completeness relation (\ref{compl}).

\end{document}